\documentstyle[12pt]{article}
\input epsf

\begin{document}

\title{\bf{Marginal Confinement in  Tokamaks by Inductive Electric Field}}

\author{P. Mart\'{\i}n, J. Puerta and E. Castro \\
\linebreak{ \ \ \ Departamento de F\'{\i}sica, Universidad Sim\'on
Bol\'{\i}var,} \\ \linebreak{\centering{ \ \ \  \ \ \ \ \ \ \ Apdo. 89000, Caracas 1080A, Venezuela.}} \\
\linebreak{ email: pmartin@usb.ve, jpuerta@usb.ve and
ecastro@usb.ve}}
\date{    }
\maketitle
\begin{abstract}
\indent \\
\indent Here diffusion and Ware pinch are analyzed as opposed
effects for plasma confinement, when instabilities are not
considered. In this work it is studied the equilibrium inductive
electric field where both effects annul each other in the sense
that the average normal velocity is zero, that is, marginal
velocity confinement is reached. The critical electric field
defined in that way is studied for different values of elliptic
elongation, Shafranov shift and triangularity. A short
complementary analysis is also performed of the variation of the
poloidal magnetic field along a magnetic line. Magnetohydrodynamic
transport theory in the collisional regime is used as in recent
publications. Axisymmetric and up-down symmetry are assumed.\\
\end{abstract}

\section{Introduction}
\indent  The H-mode is characterized by the suppression of
anomalous transport in tokamaks, because of low plasma turbulence
induces by internal barriers[1-3]. As result neoclassical
transport calculations becomes very important in this mode.
Diffusion in the collisional regime depends of the pressure
gradient at the 95\% surface adjacent to the scrape of layer or
SOL, as well as the inductive electric field $E_{\varphi}$. The
diffusion due to the gradient pressure is opposed to Ware the
pinch
effect due to $E_{\varphi}$.\\
\indent In previous papers was shown that neoclassical diffusion
can be treated with great simplicity in the case of arbitrary
plasma configuration, using a new kind of tokamaks coordinates
described there[4-6]. However, suitable direct numerical
calculations for different values of elongation, Shafranov shift
and triangularity have not been presented until now. Some previous
results on this theme using these coordinates are incomplete, and
they are not using the right parameters for the tokamaks in
operation at present.\\
\indent  Here calculations are presented in a different way, since
we look for the values of $E_{\varphi}$ in the marginal velocity
confinement, that is, when the average velocity on 95 \% surface
is zero the results corresponding to marginal confinement flux,
that is,  the transition from the outgoing to ingoing flux, could
be more interested, but much more difficult to calculate and a
suitable and simple treatment  describing this process for any
plasma configuration seem that they have not been publishing until
now.\\
\indent In the calculations now presented there are first a
suitable normalization procedure absent in previous calculations
as well as an adequate selection of tokamak parameters. The
normalization used here allows us to get results, which can be
useful for a diversity of different tokamak plasma configurations,
with different values of ellipticity, Shafranov shift and
triangularity.\\

\section{Theoretical Treatment}
\indent The collisonal transport treatment presented in previous
paper for toroidal axisymmetric plasmas can be written in a more
convenient
way using dimensionless integrals and  variables, as\\
\begin{eqnarray}
 \tilde{v} = \frac{< v >}{ -
\frac{{\eta}_{\bot}}{B^{2}_{\varphi 1}}\ (\frac{\partial
p}{\partial \sigma})_{1}} = \frac{1}{\hat{I}_{0}} [
\frac{\hat{I}_{2}}{(R_{1}/R_{c})^{2}}+
\frac{{\eta}_{\|}}{{\eta}_{\bot} {\gamma}^{2}_{1}
(R_{1}/R_{c})^{2}}(\hat{I}_{3} -
\frac{\hat{I}^{2}_{1}}{\hat{I}_{4}}) ] - \frac{\tilde{E}_{\varphi
1}}{\gamma_{1} \hat{I}_{0}} ( \hat{I}_{7} - \frac{\hat{I}_{1}
\hat{I}_{6}}{\hat{I}_{4}} + {\gamma}^{2}_{1} \hat{I}_{5}) \ \ ,
\end{eqnarray}
where $\tilde{v}$ is the dimensionless normal velocity derived
from the velocity $ < v > $ along the magnetic surface normal, and
$\tilde{E}_{\varphi 1}$ is a dimensionless electric field defined
as
\begin{eqnarray}
\tilde{E}_{\varphi} = \frac{{E}_{\varphi}}{ -
\frac{{\eta}_{\bot}}{B_{\varphi 1}}\ (\frac{\partial p}{\partial
\sigma})_{1}} \ \ .
\end{eqnarray}
Her $R_{1}$, $B_{\varphi 1}$, $B_{p 1}$, ${E}_{\varphi 1}$ and
$(\frac{\partial p}{\partial \sigma})_{1}$ are respectively the
major radius, toroidal and poloidal magnetic field, inductive
electric field and pressure gradient at the mid-plane external
point $A_{0}$ of the magnetic cross section. The plasma
resistivity are ${\eta}_{\bot}$ and ${\eta}_{\|}$ in the
directions perpendicular and parallel to the magnetic field lines
and $R_{c}$ is the minor axis radius. The dimensionless quantity
${\gamma}_{1}$ is the ratio $\gamma_{1}= B_{p1} / B_{\varphi
 1} $ between poloidal and toroidal magnetic field at point
 $A_{0}$. On the other hand the new dimensionless integrals
 $\hat{I}_{i}$, $i =1 $ to $7$, are defined as\\
\begin{eqnarray}
\hat{I}_{0} = \oint \frac{R(s) \  ds }{ R^{2}_{c} } \ \ \ ,
\end{eqnarray}
\begin{eqnarray}
\hat{I}_{1} = \oint \frac{R(s) \  ds }{ R^{2}_{c} \ \mu ( s)} \ \
\ ,
\end{eqnarray}
\begin{eqnarray}
\hat{I}_{2} = \oint \frac{R^{3}(s) \  \mu ( s) \  ds }{ R^{4}_{c}
 \ ( 1 \  + \ {\gamma}^{2}_{1} \ {\mu ( s)}^{2} ) } \ \ \ ,
\end{eqnarray}
\begin{eqnarray}
\hat{I}_{3} = \oint \frac{R^{3}(s)  \  ds }{ R^{4}_{c}
 \ ( 1 \  + \ {\gamma}^{2}_{1} \ {\mu ( s)}^{2} ) } \ \ \ ,
\end{eqnarray}
\begin{eqnarray}
\hat{I}_{4} = \oint \frac{  ds }{ R(s)
 \ ( 1 \  + \ {\gamma}^{2}_{1} \ {\mu (s)}^{2} ) \ {\mu (s)} } \ \ \ ,
\end{eqnarray}
\begin{eqnarray}
\hat{I}_{5} = \oint \frac{ {\mu (s)} \ R(s) \ ds }{R^{2}_{c}
 \ ( 1 \  + \ {\gamma}^{2}_{1} \ {\mu (s)}^{2} )   } \ \ \ ,
\end{eqnarray}
\begin{eqnarray}
\hat{I}_{6} = \oint \frac{   ds }{ {\mu (s)} \ R(s)  } \ \ \ ,
\end{eqnarray}
\begin{eqnarray}
\hat{I}_{7} = \oint \frac{ R(s) ds }{
 \ ( 1 \  + \ {\gamma}^{2}_{1} \ {\mu (s)}^{2} ) \ R^{2}_{c} \ {\mu (s)}
 } \ \ \ ,
\end{eqnarray}
where all the integrals are around a magnetic surface and $\mu
(s)$ is a function, depending of the curvature $ \kappa_{\sigma}$
of the orthogonal line family, giving by
\begin{eqnarray}
\mu (s) \ = \ exp  \ ( \  -  \ \int_{s_{A_{0}}}^{s}
\kappa_{\sigma} \ ds) \ \ \ .
\end{eqnarray}
This results are obtained using MHD equations and assuming
toroidal axisymmetry.\\
\indent In order to get numerical as well as analytic results it
is useful to express the family of magnetic cross sections by the
equations
\begin{eqnarray}
R / R_{c} = 1 +  \lambda  [ ( E - 1 ) cos \theta  + T cos ( 2
\theta )  -  \Delta ] \ \ \ ,
\end{eqnarray}
\begin{eqnarray}
z / R_{c} =   \lambda  [ ( E - 1 ) sin \theta  + T sin ( 2 \theta
)  ] \ \ \ ,
\end{eqnarray}
where $E$, $T$ and $\Delta$ are respectively ellipticity,
triangularity and Shafranov shift distortions. The parameter
$\lambda$ in this equations labels each magnetic surface. The
previous equations are a generalization of the equations presented
by Roach, et al[7]. The quantities $E$, $T$ and $\Delta$ are
dimensionless, however for the analysis and calculations are more
useful the new dimensionless quantities $\tilde{E}$, $\tilde{T}$,
$\tilde{\lambda}$, $\tilde{R}$ and $\tilde{z}$, defined
respectively as
\begin{eqnarray}
\tilde{\Delta}= \frac{\Delta}{E - 1} \ \ ; \ \ \tilde{T} = \frac{T
}{E - 1 }  \ \ ; \ \ \tilde{\lambda} = \lambda \ ( E - 1 )   \ \ ;
\ \  \tilde{R} = \frac{R}{R_{c}}   \ \ ; \ \ \tilde{z} =
\frac{z}{R_{c}} \ \ \ .
\end{eqnarray}
\indent The well know Shafranov shift $\Delta_{Shaf}$ is connected
to the previous parameters by
\begin{eqnarray}
\Delta_{Shaf} = ( \tilde{\Delta} - \tilde{T} ) \ a \ \ \ ,
\end{eqnarray}
where $ 2 a $ is the size of the 95 \% magnetic surface
measurement at the mid-plane, or in different words, $a$ is the
horizontal half-width of the plasma. The equations of the cross
section magnetic lines will be now
\begin{eqnarray}
\tilde{R}  = 1 +  \tilde{\lambda}  [ cos \theta  + \tilde{T} cos (
2 \theta )  - \tilde{\Delta} ] \ \ \ ,
\end{eqnarray}
\begin{eqnarray}
\tilde{z}  =  \frac{E + 1}{E - 1 } \ \tilde{ \lambda}  [ sin
\theta + \tilde{T} sin ( 2 \theta ) ] \ \ \ .
\end{eqnarray}
Denoting by $\lambda_{a}$ the value of the parameter $\lambda$
generating the  95 \% magnetic surface with general coordinates
$R_{a}$, $z_{a}$, then
\begin{eqnarray}
{R}_{a}  = 1 +  \tilde{\lambda}_{a}  [ cos \theta  + \tilde{T} cos
( 2 \theta )  - \tilde{\Delta} ] \ \ \ ,
\end{eqnarray}
\begin{eqnarray}
{z}_{a}  =  \left( \frac{E + 1}{E - 1 } \right) \ \tilde{
\lambda}_{a} [ sin \theta + \tilde{T} sin ( 2 \theta ) ] \ \ \ .
\end{eqnarray}
\indent The largest and smallest values of $R_{a}$ are
respectively $R_{a 1}$ and $R_{a 2}$, and its values are
\begin{eqnarray}
\tilde{R}_{a 1} = {\tilde{R}}_{a} (\theta = 0 )  = 1 +
\tilde{\lambda}_{a}  [ 1 +  \tilde{T} - \tilde{\Delta} ] \ \ \ ,
\end{eqnarray}
\begin{eqnarray}
\tilde{R}_{a 2} = {\tilde{R}}_{a} (\theta = \pi )  = 1 +
\tilde{\lambda}_{a}  [ - 1 +  \tilde{T} - \tilde{\Delta} ] \ \ \ .
\end{eqnarray}
\indent The radius $\tilde{R}_{a 0}$ of the center of the plasma
and plasma size $a$ will be respectively
\begin{eqnarray}
\tilde{R}_{a 0} =  \frac{\tilde{R}_{a 1} \ + \  \tilde{R}_{a
0}}{2}= 1 + \tilde{\lambda}_{a} [  1 + \tilde{T} - \tilde{\Delta}
] \ \ \ ,
\end{eqnarray}
\begin{eqnarray}
\frac{a}{R_{c}} =  \frac{\tilde{R}_{a 1} \ - \  \tilde{R}_{a
0}}{2}=  \tilde{\lambda}_{a} \ \ \ .
\end{eqnarray}
\indent It seems also convenient to connect that parameters with
the aspect ratio $A$, such that
\begin{eqnarray}
A = \frac{{R}_{a 0}}{a} = \frac{( 1 + \tilde{\lambda}_{a}
(\tilde{T} - \tilde{\Delta} )}{ \tilde{\lambda}_{a}} \ \ ; \ \
 \frac{1}{ \tilde{\lambda}_{a}}  = A + \tilde{\Delta}
- \tilde{T} \ \ \ .
\end{eqnarray}
\indent As in our previous papers, $b$ is the maximum value of
$z_{a}$. The elliptic elongation $K$ and triangularity are
connected to our previous parameters as
\begin{eqnarray}
K = b / a =  \frac{E + 1}{E - 1 } \left( \frac{3}{4} + \frac{1}{4}
\sqrt{ 1 + \tilde{g} } \right)  \sqrt{1 + \tilde{h}} \ \ \ ,
\end{eqnarray}
where $\tilde{g}$ and $\tilde{h}$ are
\begin{eqnarray}
\tilde{g} =  32 \ \tilde{T}^{2} \left( \frac{E + 1}{E - 1 }
\right)^{2} \ \ \ ; \ \ \ \tilde{h} = \frac{1}{\tilde{g}} \ \left(
\sqrt{1 + \tilde{g}} - 1 \right)
\end{eqnarray}
and
\begin{eqnarray}
\delta =  \frac{{\tilde{R}}_{a 0} - {\tilde{R}}_{a m}}{(a /
R_{c})} = \left(  \frac{\tilde{T}}{ E + 1 )}\right) \left[ 2 \
\tilde{h}  \ ( E - 3 ) - ( E + 1 ) \right] \ \ \ ,
\end{eqnarray}
where $R_{a m}$ is the radius of the point with maximum z.\\
\section{Results }
\indent The dimensionless variables defined previously simplifies
the computation, because quantities we need for the calculation
are:  the ratio between poloidal and toroidal magnetic fields
$\gamma_{1}$, the horizontal half-width of the plasma $a$, the
aspect
 ratio $A$, the ratio ${\eta}_{\parallel}/{\eta}_{\perp}$, the
 Shafranov shift, the ellipticity $K$ and triangularity $\delta$.
 From $K$ and  $\delta$, the values of $E$ and $\tilde{T}$ are
 determined. The Shafranov shift, the value of $a$ and the
 calculated value $\tilde{T}$ together allow the calculation of  $\tilde{\Delta}$ and
 $\tilde{\lambda}_{a}$.\\
 \indent In this way the family of magnetic surfaces can be
 determined. Following the procedures described in previous works
 the family of orthogonal lines can be also determined, which
 allows to obtain the curvature function $\kappa_{\sigma}$ and
 the function $\mu(s)$. All the integrals needed to obtain the
 normal dimensionless velocity $\tilde{v}$ can also be performed
 once the value of $\gamma_{1}$ is given. The velocity $\tilde{v}$
 can also be obtained if the ratio
 ${\eta}_{\|}/{\eta}_{\bot}$ is given,  and it is  a linear function
 of the dimensionless toroidal electric field $\tilde{E}_{\varphi
 1}$. The intersection of that value with the axis of abscissa
 allows to determine the critical dimensionless electric field $\tilde{E}_{\varphi
 1 crit}$, for marginal  velocity confinement.
 This field will be later determine for different values of
 ellipticity $K$, dimensionless Shafranov shift
 $\Delta_{Shafranov}/a$, and triangularity $\delta$. \\
 \indent Following the above described procedure, the ellipticity
 $K$ and triangularity $\delta$, in Figure 1,  are given as $ K = 1.76 $ and $ \delta = 0.25
 $.  These correspond to values of $E$ and $\tilde{T}$: $\tilde{E} = 4
 $, $\tilde{T} = 0.3 $. The real procedure we use was a little
 different. We first  select values of $E$ and $\tilde{T}$, in such way $K$
 and $\delta$ become about their values in JET tokamak[8].
 This procedure is simpler for us, and it is the same idea.\\
 \indent In Figure 1, several cross sections magnetic lines have
 been drawn for different values of $\lambda$ in the interval
 from
 zero to $\lambda_{a}$, giving in  Eq.(24). The characteristic values for
 that figure are: elipticity $K = 1.76$; relative Shafranov shift
 ${\tilde{\Delta}}_{Shaf} = {\Delta}_{Shaf} / a = 0.3 $;
 triangularity $ \delta = 0.25$; horizontal half width $a = 1.12 \ m$
 ( 95 \% surface ); aspect ratio $A = 2.5 $ and  minor magnetic axis
 ratio $R_{c} = 3 \ m$. From the previous data, the following
 parameters are determined: elliptic dispersion $E = 4 $, see Eq.(25)to (27), relative
 triangularity dispersion $ \tilde{T} = T / ( E -1) = 0.3 $;
 relative Shafranov dispersion ${\tilde{\Delta}} = {\tilde{\Delta}}_{Shaf} + \tilde{T} =
 0.9$; relative $\lambda$-parameter at the 95 \% surface
 $\tilde{\lambda}_{a} = ( A + \Delta - T)^{-1} =
 0.1075$;  the outward radius $ R_{A_{0}}$,  $R1 = R(\theta = 0) = R_{A_{0}}=
 R_{c} [ 1 + \tilde{\lambda}_{a}(1 + \tilde{T} - \tilde{\Delta})] =
 3.38 \ m  $; the inward radius $R2 = R(\theta = \pi) =
 R_{c} [- 1 + \tilde{\lambda}_{a}(1 + \tilde{T} - \tilde{\Delta})] =
 1.45 \ m$; center plasma radius $R_{0} = (R_{1} + R_{2})/2 = 2.42 \
 m$ and radius at the maximum z, $((dz/dR)_{R_{m}} = 0)$, $R_{m} = 2.18 \
 m$.\\
 \indent In Figure 2, the function $\mu(s)$ is show as a function
 of $\theta$. This function allows to determine poloidal field
 around a magnetic surface, once the value $B_{p 1}$ at the
 outward point $A_{0}$ is measured or determined. In order to find
 $\mu(\theta)$, it is necessary to determine the curvature of the
 family of orthogonal lines. The procedure has been explained
 elsewhere[4]. The function $\mu(\theta)$ appears in most of the
 integrals needed to determine the average normal velocity $ < v > \ $
 first calculated by Pfirsch-Schl\"{u}ter for  cross-sections of
 circular magnetic surfaces[9]. The function to be used in this
 work is the central function, plain line. However, two other
 $\mu$-functions are also shown to illustrate that function. The
 upper curve correspond to the case where the triangularity and
 Shafranov shift are zero. The lower curve is just the case of
 zero triangularity, but the same Shafranov shift than in the main
 curve ( central one), where the triangularity is $\delta = 0.25$,
 as in Figure 1.\\
 \indent Since the function $\mu(s)$ shown essentially the behavior
 of $R B_{p}$, the upper curve illustrate that this product is
 constant at the inward and outward points, when there is not
 triangularity and Shafranov shift. However, the value of $R
 B_{p}$, decreases at the uppest and lowest points because of the
 ellipticity elongation $K = 1.76$. The Shafranov shift modifies
 this pattern in the way that the values at the inward area are almost
 constant, and very low compared with those at the outward point
 $A_{0}$. Introducing triangularity as in the case of the central
 line does not modify the pattern, but the values at the flat part
 of the curve are not so low.\\
 \indent In Figure 3, the dimensionless average normal velocity is
 shown as a function of the dimensionless toroidal electric field
 $\tilde{E}_{\varphi 1}$ at point $A_{0}$, the line here found is the
 straight line, because of the normalization here used using the
 value of $(\eta_{\perp}/ B_{\varphi} )(\partial p /\partial
 \sigma)_{1}$  at the point $A_{0}$. The intersection of that line
 with the abscissa define the critical dimensionless toroidal
 field  $\tilde{E}_{\varphi 1}$ for marginal velocity
 confinement. This critical electric field is show as a function
 of the elliptic elongation $K$, where all the other variables are
 kept fixed at the values shown in Figure 1, that is, $\delta =
 0.25 $ and $\tilde{\Delta} = 0.9 $. To determine the integrals in
 order to draw Figure 3 and 4, the value of $\gamma_{1}$ is
 chosen as 0.3, which correspond for instance to toroidal magnetic
 field $B_{\varphi 1} = 5 \  Teslas$ and poloidal magnetic field $B_{p 1} = 1.5
 \ Teslas$. The ratio of perpendicular to parallel resistivity,
 ${\eta}_{\parallel}/{\eta}_{\perp}$ has been considered as 1.97
 as in page 669 of Ref.(8). As a way of completion $T_{e}$ is taken
 as $5 \ keV$.\\
 \indent The Figure 3 shown that the outward velocity due to
 diffusion is opposite by the inward effect due to the inductive
 electric fields $E_{\varphi}$. After a critical value $\tilde{E}_{\varphi
 }$, the characteristic Ware pinch effect[10] becomes more important, and an average
 inward velocity $ < \tilde{v} > \ $  is produced, in such a way
 that the plasma appears confined as long as instabilities are not
 considered. The critical dimensionless electric field increase
 with the elliptic elongation $K$ if all the parameters are kept
 fixed, however, the variation is not so large, from 0.8 a factor to about
 1.1 as it is shown in Figure 4. In this figure  has been normalized
  with a second procedure. First a normalized critical  toroidal electric
  field is selected as reference and denoted by $\tilde{E}_{\varphi 1 \ crit. \
  ref.}$, which in this case corresponds to that given in Fig. 3,
  where the characteristic values of the parameters above given.
  The value of this critical dimensionless toroidal electric field is
  $29.8106$, and correspond also to the horizontal line through
  one, which it is also show by a dot line. This kind of
  normalization is also performed in Figure 5 and 6. The toroidal
  electric fields with the second normalization explain above are
  denoted by ${\hat{E}}_{\varphi 1 \ crit.}$.\\
 \indent The changes due to Shafranov
 shift with all the other parameters fixed, are more significative
 as illustrates the Figure 5, where the critical electric field
 could be one forth or 3 times the value of that shown in Figure
 3. Furthermore the curve in Figure 4 is almost linear, but not
 that in Figure 5, which seems somewhat as a parabola. Finally, in
 Figure 6, the change of $\tilde{E}_{\varphi } crit$ with
 triangularity are shown. There also  ${\hat{E}}_{\varphi 1 \ crit.}$
 changes strongly with $\delta$. More important than this it is that
 the changes are very significative for low values of $\delta$,
 and ${\hat{E}}_{\varphi 1 \ crit.}$ could be lower by a factor 5, and with values of
 triangularity no so large, as $\delta = 0.25 $. Here also the
 curve is strongly not linear, however the concavity of the curve is
 opposite to that in Figure 6.\\
\section{Conclusion}
\indent In most tokamak operation the inductive magnetic field
effect exceeds the plasma diffusion effect and the Ware pinch
effect contract the toroidal plasma column. Here the critical
point where both effect becomes almost equals is studied. This
equilibrium situation is consider as that where the average normal
velocity becomes null or void, which will be defined as the
marginal velocity confinement. A suitable normalization procedure
allows to extend our analysis to a large amount of different
situations in tokamak plasma configurations. The critical toroidal
electric field changes very little with the ellipticity of the
plasma. However, the changes are very strong with the Shafranov
shift and triangularity. The curves in those last cases seems
parabolas, but with opposite curvature. Very large changes for
small values of triangularity have been found, producing changes
with a factor 5
for modest triangularity values as $\delta = 0.25$.\\

\bigskip\parindent=0pt{\bf References}\smallbreak

[1]  K. H. Burrell, Phys. Plasmas {\bf 4}, 1499 ( 1997).

[2] V. B. Lebeder and P. H. Diamond, Phys. Plasmas {\bf 4}, 1087
(1997)

[3] P. H. Diamond, Y. M. Liang B. A. Carreras and P. W. Terry,
Phys. Rev. Letters {\bf 72}, 2565 (1994)

[4] P. Martin and M. G. Haines, Phys. Plasmas {\bf 5 } 410 (1998).

[5] P. Martin, Phys. Plasmas {\bf 7}, 2915 ( 2000).

[6] P. Martin and J. Puerta, Physica Scripta {\bf T-84}, 212
(2000).

[7] C. M. Roach, J. W. Connor and S. Janjua, Plasma Phys. Control
Fusion {\bf 37}, 679 (1995)

[8]  John Wesson, " Tokamaks " ( Clarendon Pres-Oxford, 1997, 2nd
Edition ), pp. 555,669.

[9] D. Pfirsch and A. Schl\"{u}ter, Max-Planck Institute Report
MPI/PA/7/62 ( 1962)

[10] A. A. Ware, Phys. Rev. Letters {\bf 25 }, 15 ( 1970)

\begin{figure}
\begin{center}
\epsfxsize=10.0 cm \epsfysize=18 cm \epsfbox{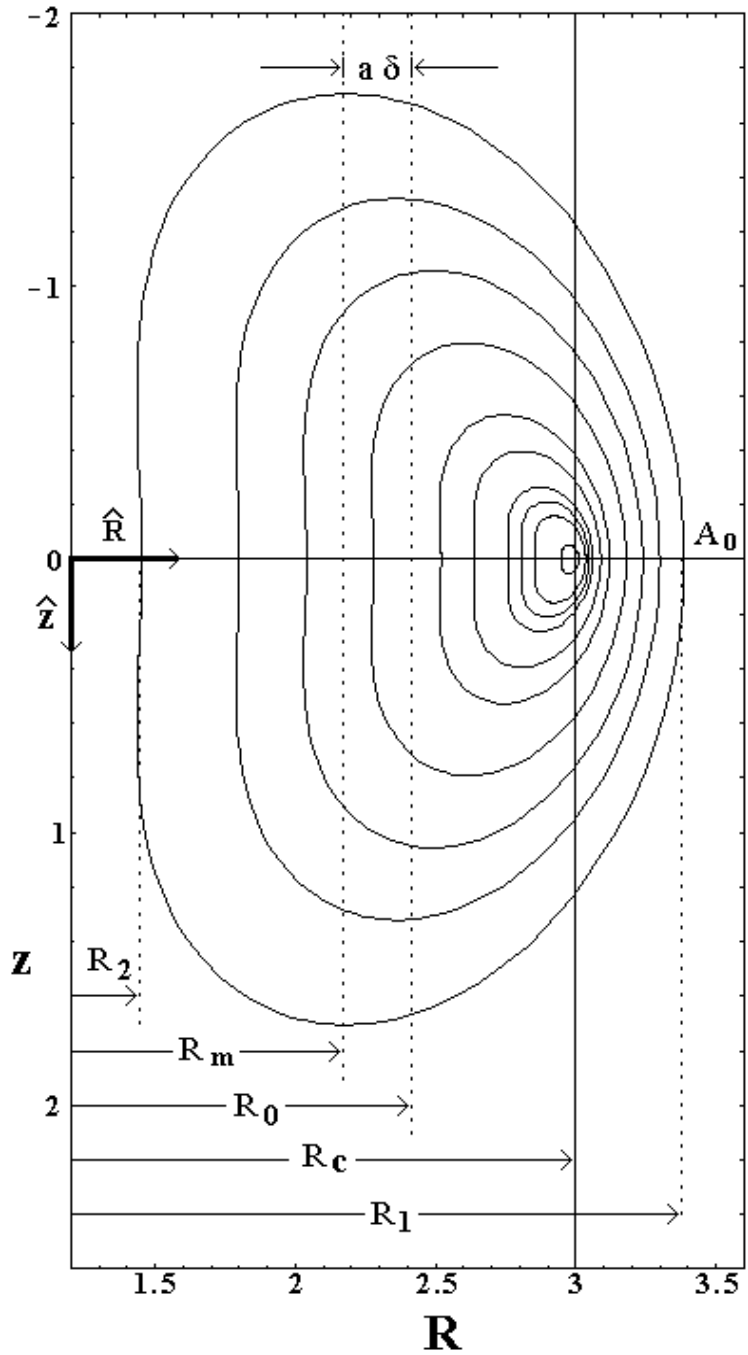}
\end{center}
\caption{\label{label} Magnetic flux surfaces showing the
characteristic radius, triangularity and main point, with the
parameters values given in the text.}
\end{figure}

\begin{figure}
\begin{center}
\epsfxsize=14.0 cm \epsfysize=14 cm \epsfbox{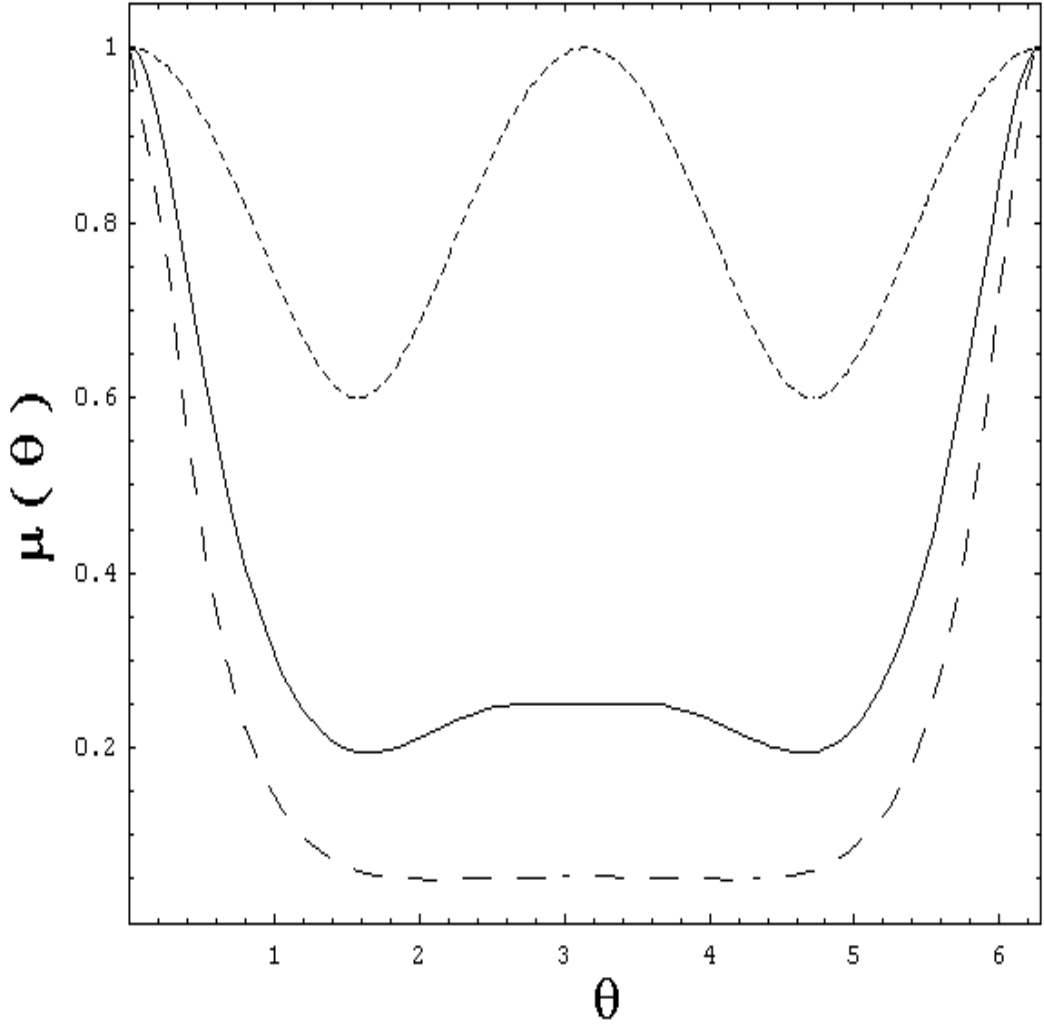}
\end{center}
\caption{\label{label} Characteristic exponential factor around a
magnetic surface for the main values of the reference surfaces,
plain line, and two other situations. No triangularity but
Shafranov shift, dashed line, and neither Shafranov  and
triangularity, point line. The parameters values of the reference
curve are: $E = 4$, $\tilde{{\Delta}}= 0.9 $, $ A = 2.5 $, $
\tilde{T}  = 0.3 $, $\gamma_{1} = 0.3$ and $R_{c} = 3 \ m$}
\end{figure}

\begin{figure}
\begin{center}
\epsfxsize=14.0 cm \epsfysize=14 cm \epsfbox{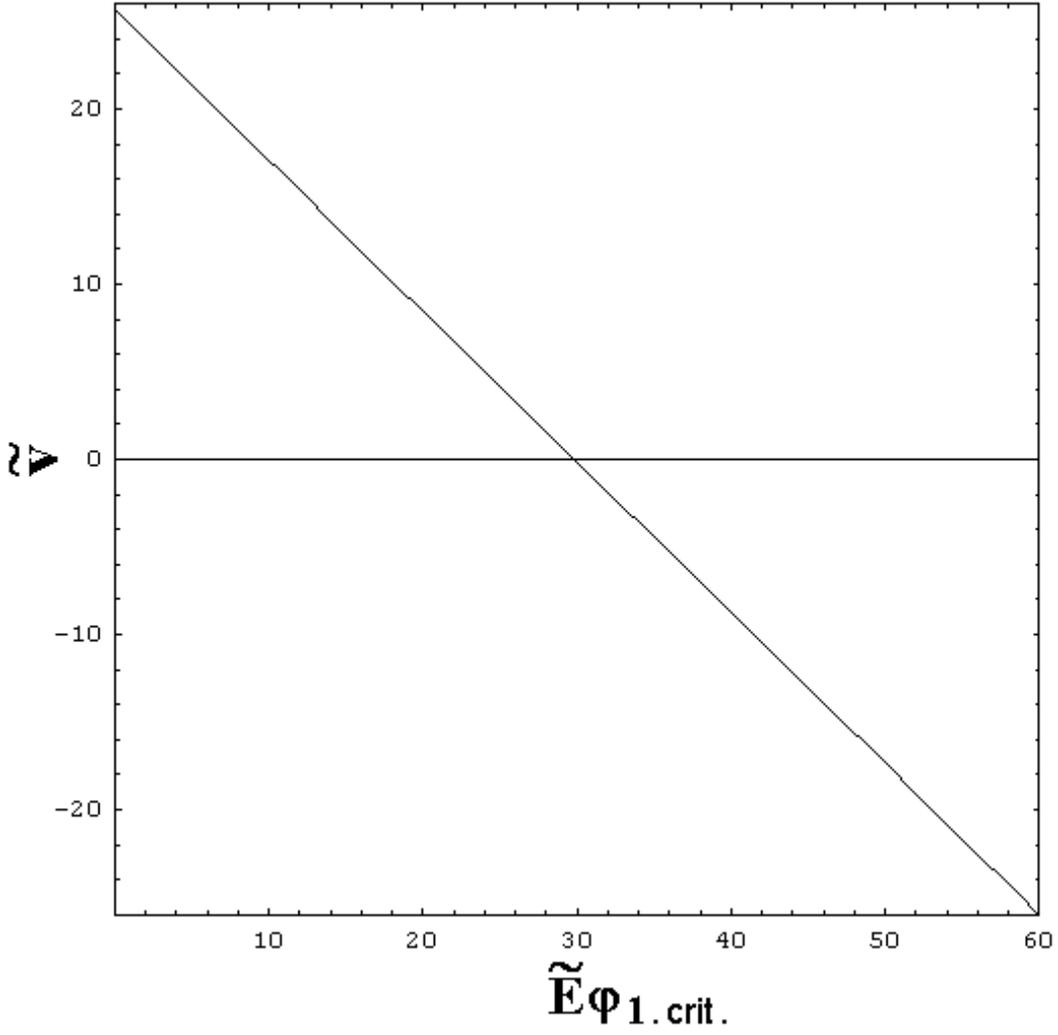}
\end{center}
\caption{\label{label} Normalized velocity versus normalized
toroidal electric field in the outward point for the parameters
values described as reference parameters in the text and for the
outward.}
\end{figure}

\begin{figure}
\begin{center}
\epsfxsize=14.0 cm \epsfysize=14 cm \epsfbox{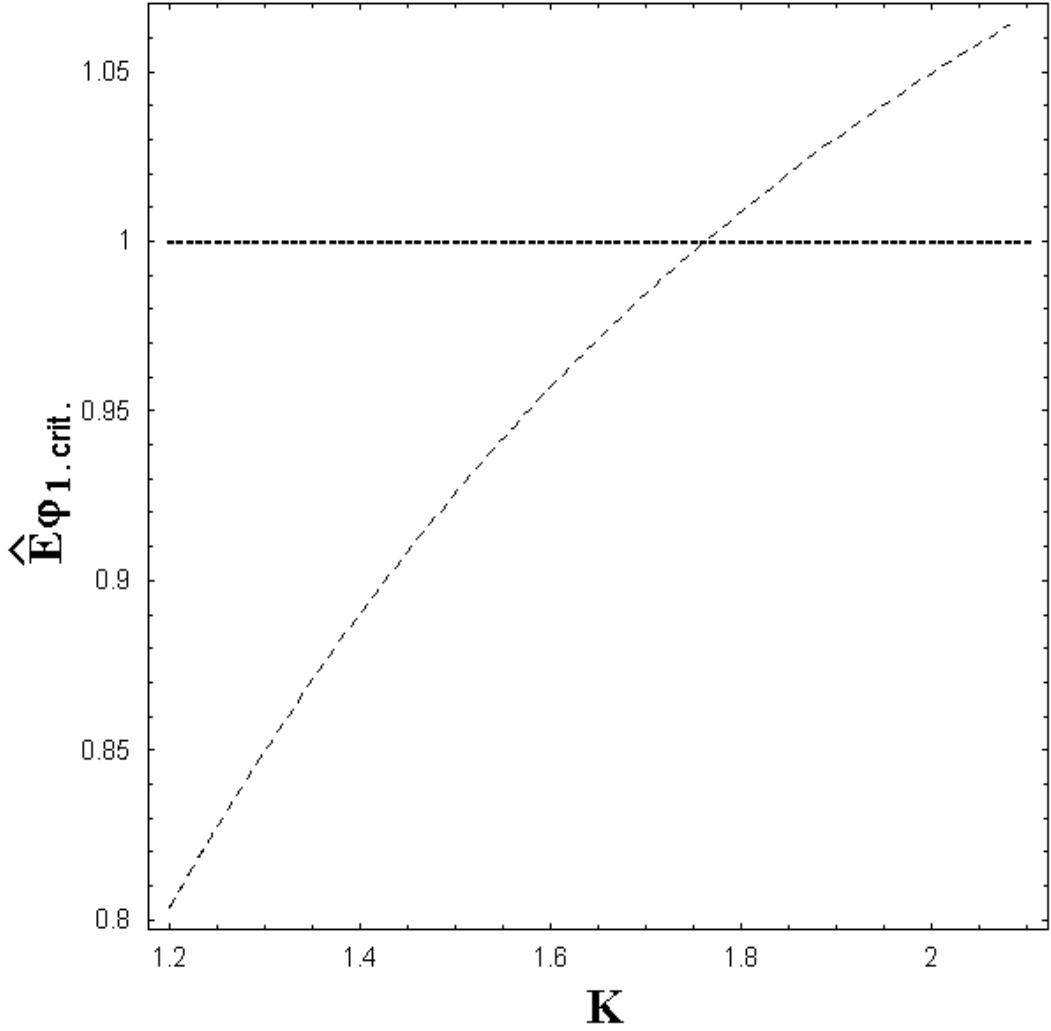}
\end{center}
\caption{\label{label} Twofold normalized critical electric field
${\hat{E}}_{\varphi 1 \ crit.}$ as a function of elliptic
elongation with the triangularity and Shafranov shift equal to
those in the reference curves, Fig. 1 and Fig. 3 }
\end{figure}

\begin{figure}
\begin{center}
\epsfxsize=14.0 cm \epsfysize=14 cm \epsfbox{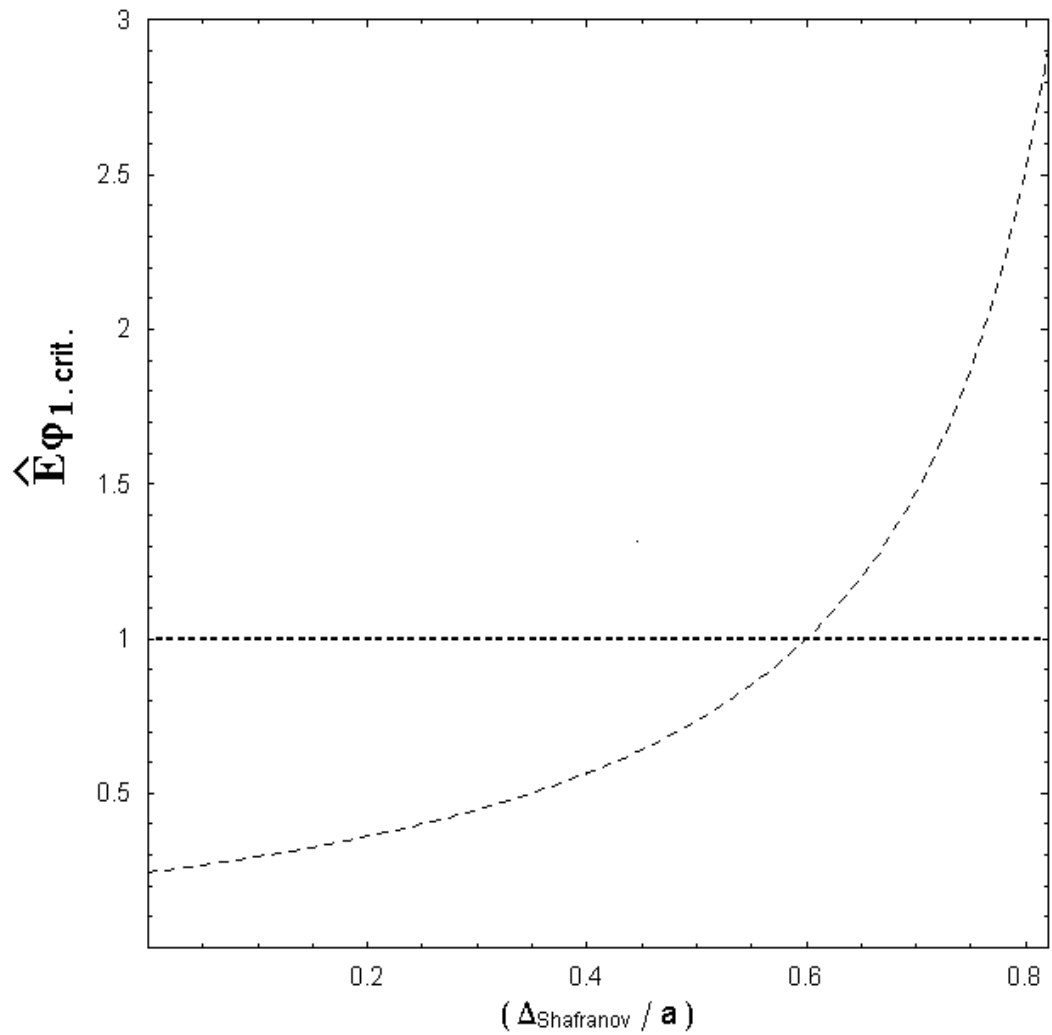}
\end{center}
\caption{\label{label} Twofold normalized critical electric field
${\hat{E}}_{\varphi 1 \ crit.}$ as a function of dimensionless
Shafranov shift with elliptic elongation and Shafranov shift given
by the reference curve }
\end{figure}

\begin{figure}
\begin{center}
\epsfxsize=14.0 cm \epsfysize=14 cm \epsfbox{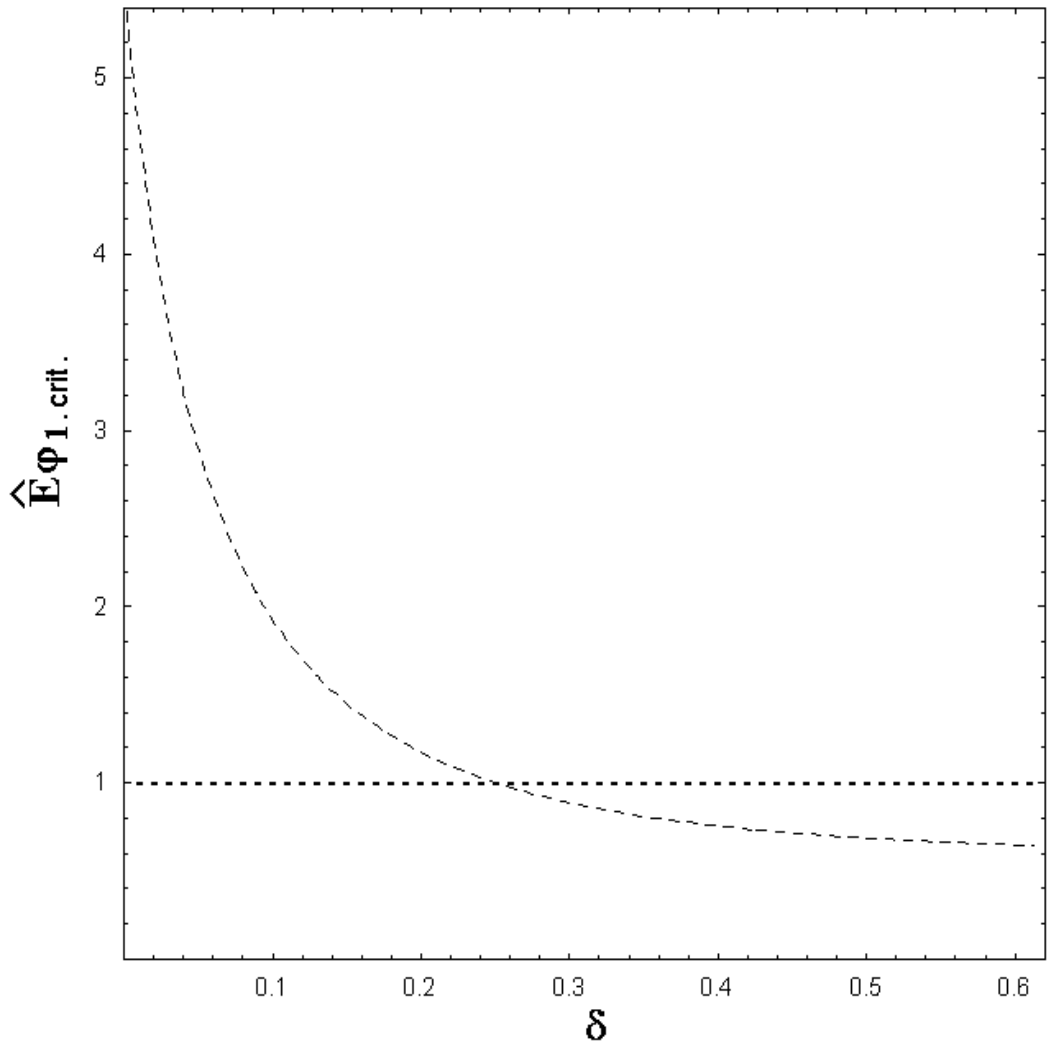}
\end{center}
\caption{\label{label}Twofold normalized critical electric field
${\hat{E}}_{\varphi 1 \ crit.}$ as a function of triangularity
with ellipticity elongation and Shafranov shift given by the
reference curve. }
\end{figure}

\end{document}